# Know Your Enemy: Characteristics of Cyber-Attacks on Medical Imaging Devices

Tom Mahler[1,a], Nir Nissim[2,b], Erez Shalom[1,c],
Israel Goldenberg[3,d], Guy Hassman[3,e], Arnon Makori[3,f], Itzik Kochav[3,g], Yuval Elovici[1,h], Yuval Shahar[1,i]

[1]Department of Software and Information Systems Engineering,
Ben-Gurion University of the Negev, Beer-Sheva, Israel
[a]tommahler@gmail.com, [c]erezsh@post.bgu.ac.il, [h]elovici@inter.net.il, [i]yshahar@bgu.ac.il

[1]Malware-Lab, Cyber-Security Research Center,
Ben-Gurion University of the Negev, Beer-Sheva, Israel
[b]nnirni.n@gmail.com

[3]Clalit Health Services, Tel-Aviv, Israel
[d]israel@clalit.org.il, [e]guyha3@clalit.org.il, [f]arnonma@clalit.org.il, [g]kochav@clalit.org.il

**ABSTRACT**

**Purpose:** Used extensively in the diagnosis, treatment, and prevention of disease, *Medical Imaging Devices* (MIDs), such as *Magnetic Resonance Imaging* (MRI) or *Computed Tomography* (CT) machines, play an important role in medicine today. MIDs are increasingly connected to hospital networks, making them vulnerable to sophisticated cyber-attacks targeting the devices' infrastructure and components, which can disrupt digital patient records, and potentially jeopardize patients' health. Attacks on MIDs are likely to increase, as attackers' skills improve and the number of unpatched devices with known vulnerabilities that can be easily exploited grows. Attackers may also block access to MIDs or disable them, as part of ransomware attacks, which have been shown to be successful against hospitals.

**Method and Materials:** We conducted a comprehensive risk analysis survey, based on the *Confidentiality, Integrity, and Availability* (CIA) model, in collaboration with our country's largest health maintenance organization, to define the characteristics of cyber-attacks on MIDs. The survey includes a range of vulnerabilities and potential attacks aimed at MIDs, medical and imaging information systems, and medical protocols and standards such as DICOM and HL7.

**Results:** Based on our survey, we found that CT devices face the greatest risk of cyber-attack, due to their pivotal role in acute care imaging. Thus, we identified several possible attack vectors that target the infrastructure and functionality of CT devices, which can cause: 1. Disruption of the parameters' values used in the scanning protocols within the CT device (e.g., tampering with the radiation exposure levels); 2. Mechanical disruption of the CT device (e.g., changing the pitch); 3. Disruption of the tomography scan signals constructing the digital images; and 4. Denial-of-Service attacks against the CT device.

**Conclusion:** Cyber-attacks on MIDs will become a major challenge to device manufacturers and healthcare providers. To ensure a safe healthcare environment and protect patients, users must be aware of the risks and understand the mechanisms behind these potential attacks.

**Clinical Relevance/Application:** New approaches for detection and prevention, such as we propose to develop, should be deployed and implemented.

## I. INTRODUCTION

In the last few decades, the healthcare industry has begun to adopt the use of computers for clinical use. Computers enhanced medical imaging by adding novel and innovative capabilities, enabling early discovery of diseases, research of new diseases, better treatment of medical conditions, etc. These computational capabilities led to accelerated research and development in this field, and CT and MRI quickly became an integral part of many medical treatment and procedures. In 2000, imaging was ranked as one of the top medical developments in the past 1000 years, by the *New England Journal of Medicine* [1].

This trend of nonlinear growth in use of CT and MRI is supported by data we collected from the *Organization for Economic Co-operation and Development* (OECD) *Health Statistics database* [2] for number of CTs and MRIs and medical exams (at hospital and at ambulatory care) from the last 30 years: The total growth of number of CTs and MRIs in the first 25 years, 1980-2005, is similar to the total growth in the last 10 years, 2005-2014 (see Fig. 1), and the total growth of number of exams in the first 17 years, 1990-2007, is similar to the total growth in the last 7 years, 2007-2014 (see Fig. 2). By using polynomial regression (by the order of 2) we predict that by 2020 the current number will increase by 40%, nearly reaching 2000.

While this development, usually, improves the overall treatment that patients receive, it also introduces MIDs to new cyber-threats. Today's CTs and MRIs are more digital and more connected, making these devices vulnerable to network related cyber-threats. CTs and MRIs are used all the time, and for various reasons such as supporting life-saving treatments. This makes CTs and MRIs a critical resource in the hospital. Moreover, these devices are very expensive, thus, very few of them, if at all, are held by hospitals. Therefore, failure of one device may sabotage the entire hospital's operation, influencing many people.

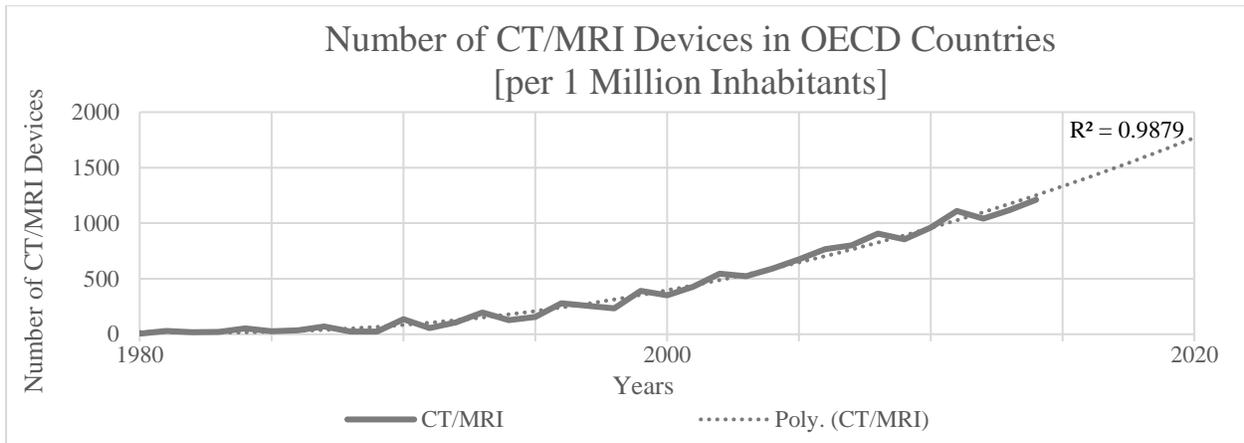

Fig. 1. Number of CT/MRI devices in OECD countries in years 1980-2014 per 1 Million inhabitants, and an extrapolation performed by using a second-order regression line.

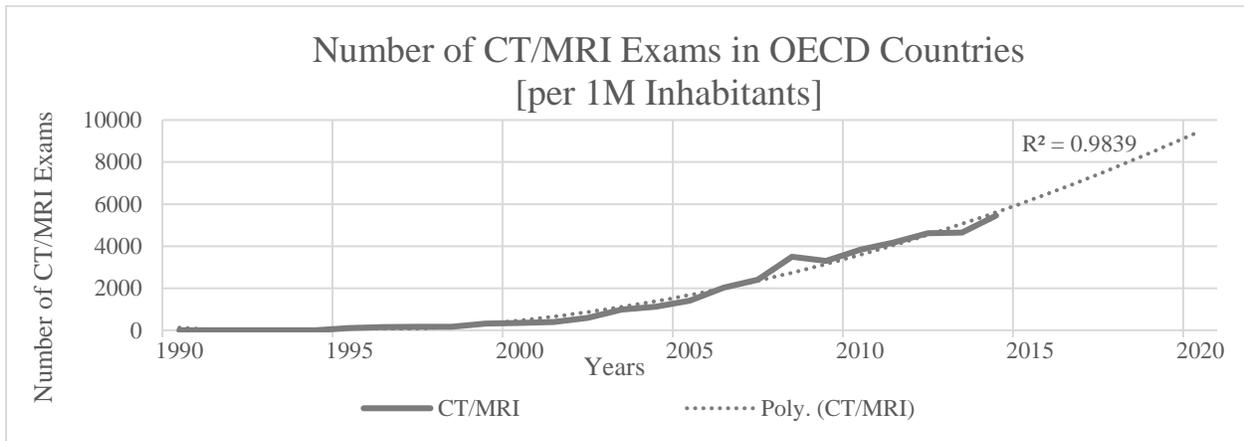

Fig. 2. Number of CT/MRI exams in OECD countries in years 1990-2014 per 1 Million inhabitant, and an extrapolation performed by a second-order regression line.

In Fig. 3, we can see a graph of new malware discovered each year, based on data from AV-TEST [3]. Today, each year more than 120 Million new malware are discovered. In order to keep up with this extremely fast development of cyber-threats, CTs and MRIs manufacturers must adapt quickly to new threats. In general, CTs and MRIs are not cyber-security oriented. Many medical devices development process takes years. It is estimated that time from concept to market for medical devices is 3-7 years [4]. This complex process requires not only state-of-the-art technology, but also many regulations and approval, and for good reason - people life is at stake, which takes a significant amount of time. Due to these strict regulations, by the time CTs and MRIs manage to adapt, the cyber-threats are already completely different, making CTs and MRIs very vulnerable.

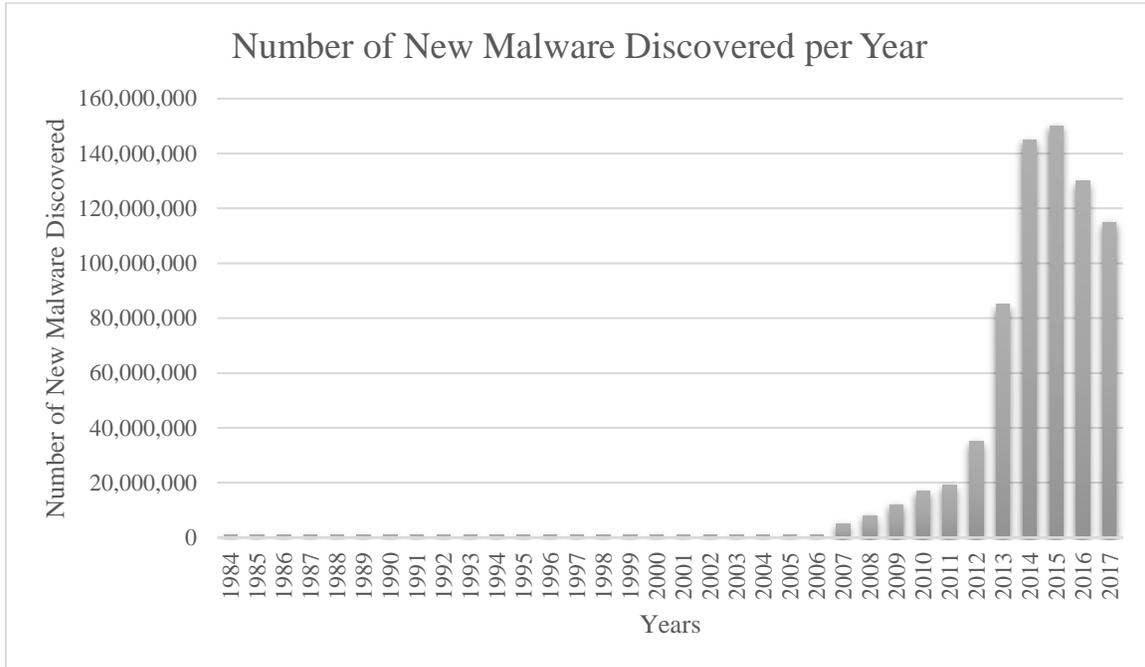

Fig. 3. A graph of new malware discovered each year, based on data from AV-TEST [3]. Taken at 18.12.2017.

## II. CASE-STUDY: THE WANNACRY CYBER-ATTACK

The WannaCry [5], [6], [7] is a world-wide cyber-attack that was held in May 2017, reported by many top-leading sites. We can see a screenshot from with the actual ransom note from the WannaCry attack in Fig. 4. The attack was done using the WannaCry crypto-warm, and reports estimate that it infected over 200,000 devices in more than 150 nations [8], including a major impact on the U.S. delivery company FedEx, Spanish telecoms and gas companies, French Renault car production factories, Russian interior ministry, and the U.K. *National Health Service* (NHS) [9]. This was the first, large scale cyber-attack that affected healthcare industry directly, by infecting tens of Thousands of the NHS's hospitals' devices, including MIDs such as MRI (according to *The Sunday Times* reports [10]), causing them to be nonoperational by encrypting them. This attack caused several hospitals to turn down patients [11], [12] and divert ambulances routes [13].

In April 2017, the hacker group *The Shadow Brokers* leaked the *EternalBlue* exploit tool, among several other exploits, believed to belong to the NSA. The *EternalBlue* (CVE-2017-0144) [14] is an exploit of Windows' *Server Message Block* (SMB) protocol which affected many Windows versions such as: Windows Vista, Windows Server 2008, Windows 7, Windows 8.1, Windows Server 2012, Windows RT 8.1, Windows 10 and Windows Server 2016. The exploit allows arbitrary code execution via crafted packets by remote attackers.

Even though in March 2017, before the exploit was leaked, Microsoft issued critical security patch that fixed the *EternalBlue* exploit [14], many computers which are not regularly updated (due to lack of care, or to technical issues) were still vulnerable. In particular, this includes the computers of the hospitals in the U.K., and medical devices, since the devices were not patched fast enough.

**WannaCry's Kill Switch [15]:** WannaCry used special URL, to track activity from infected machines. This domain was found to act as a "kill switch" that shut down the software before it executed its payload, stopping the spread of the ransomware. Some believe that this was used to prevent it from running on quarantined machines used by anti-virus researchers [16]. Some sandbox environments respond to all queries with traffic in order to trick the software into thinking that it is still connected to the internet. WannaCry attempts to contact an address which did not exist, to detect whether it was running in a sandbox, and if so, do nothing. When discovered, some hackers even tried using a *Mirai* botnet variant to create a distributed attack on WannaCry's kill-switch domain with the intention of knocking it offline and reigniting the attack [17].

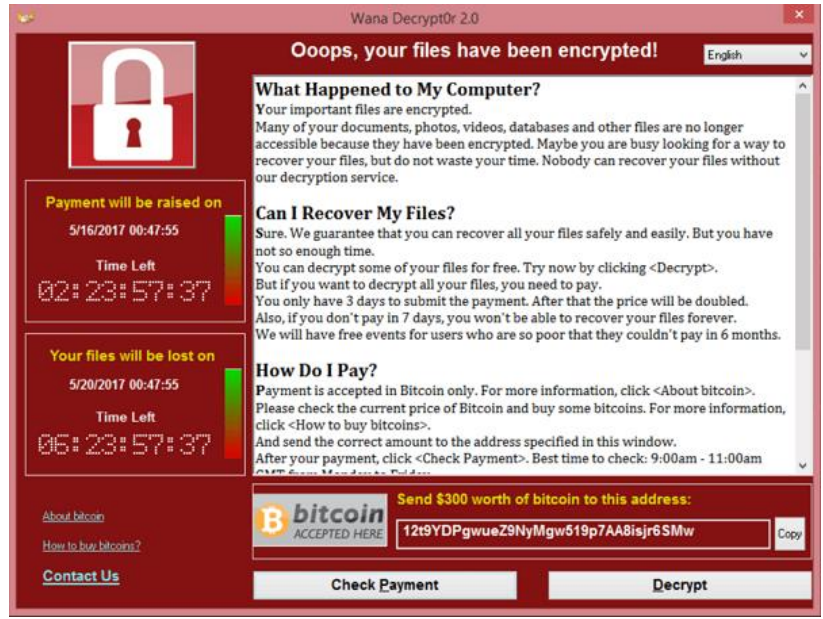

Fig. 4. A Screenshot of the ransom note left on an infected system [18].

### III. CYBER-ATTACKS ON A CT DEVICE

CT devices use ionizing radiation to perform CT exam. This, along with its high price and criticality to the hospital, makes cyber-attacks on it very dangerous. In this paper, we list several cyber-attacks that we have conducted, which target CT devices.

*1. Configuration Files Disruption*

The CT entire scan operation is defined in the scan configuration file, inside the host control PC, making it extremely important. By manipulating the file, the attack is able to change the CT's behavior. The impact of this attack is that the malware can control the entire CT operation, putting the patient in a very high risk. Some implementations are presented in the next attacks.

*2. Mechanical Disruption of MID's Motors*

MIDs' ecosystem includes several components with mechanical motors, being controlled by the MID: mechanical bed, scanner motors, rotation motors, etc. The motors receive instructions from a control unit (e.g., the host control PC). This makes possible to change the commands controlling these motors, causing undesired movement of these motors. The impacts of this attack can be direct physical damage to the device by misusing the motors, and potential physical damage to the patient by making the motors move in a way that hurts the patient.

*3. Disruption of Image Results*

The CT's ecosystem includes an image reconstruction component which is responsible of reconstructing the image based on the raw data. In addition, this component relates to a specific medical record of the patient being scanned to the image results. The results, including the relation to the patient, are being sent via the host control PC using DICOM protocol. This component can be the host control PC, or another computer for this task. This attack disrupts the exam's results, requiring a second exam. A more sophisticated attack may, instead, alter the results which can be much more dangerous, since, if performed well, it is very difficult to tell that something is wrong. An even more sophisticated attack can disrupt this relation between the results and the patient, by either relating the result to the wrong patient, or relating wrong results to the patient being scanned. The impacts of these attacks may be severe in cases which require immediate care to the patient which even the smallest delay may be fatal. In addition, by either removing dangerous tumors from the image, or adding non-existing tumor to the image, the attack can lead to dangerous mistreatment. These scenarios could be fatal cases.

*4. Ransomware DoS*

A ransomware is a type of malware, which encrypts the victim's files with a strong encryption, demanding a ransom in order to decrypt the files. Since MIDs rely on computers to operate, encrypting the host control PC will make the MID nonoperational. In addition, encrypting scan results in critical situations may put patients at risk. The WannaCry attack mentioned previous is an example of a ransomware attack.

## IV. CONCLUSION

Cyber-attacks on MIDs will become a major challenge to device manufacturers and healthcare providers. We've seen an example for this, by examining the case study of the WannaCry attack to learn how MIDs are vulnerable. In order to ensure a safe healthcare environment and protect patients, users must be aware of the risks and understand the mechanisms behind these potential attacks.

From the results of our survey, we learn that the host control PC is the most vulnerable component in the CT's ecosystem, due to is central position and controls. Therefore, it seems that protecting it from cyber-attacks can significantly increase the overall security of CT device. As we've seen, due to regulations it is hard to maintain regular updates to this component. Therefore, common techniques for securing a computer, such as installing anti-virus protection, are insufficient for the prevention of cyber-attacks.

Therefore, other methods must be researched in order to better protect these devices against cyber-attacks. In our next papers, we will present a novel technique for securing CT devices, based on machine learning. Our technique will take an out-of-band approach, by learning the actual commands being sent to the CT's gantry, together with patients' profiles and scan labels and detecting anomalies. Such method can assume that the host control PC is infected, and by taking an out-of-band approach, we can examine the sent commands before they arrive to the CT's gantry, enabling to prevent malicious commands before they occur.